\documentclass{aastex6}

\begin{document}


\title{On the non-Kolmogorov nature of flare-productive solar active regions}


\author{Revati S. Mandage\altaffilmark{1} and R.~T.~James McAteer\altaffilmark{2}}







\altaffiltext{1}{Physics and Astronomy Department, Rice University, 6100 Main MS-61, Houston, TX 77005-1827 USA}
\altaffiltext{2}{mcateer@nmsu.edu Department of Astronomy, New Mexico State University, MSC 4500, Las Cruces, NM 88001 USA}


\begin{abstract}

 A magnetic power spectral analysis is performed on 53 solar active regions, observed from August 2011 to July 2012. Magnetic field data obtained from the Helioseismic and Magnetic Imager, inverted as Active Region Patches, are used to study the evolution of the magnetic power index as each region rotates across the solar disk. Active regions are classified based on the number, and sizes, of solar flares they produce, in order to study the relationship between flare productivity and the magnetic power index. The choice of window size and inertial range plays a key role in determining the correct magnetic power index. The overall distribution of magnetic power indices has a range of $1.0-2.5$. Flare-quiet regions peak at a value of 1.6, however flare-productive regions peak at a value of 2.2. Overall, the histogram of the distribution of power indices of flare-productive active regions is well separated from flare-quiet active regions. Only 12\% of flare-quiet regions exhibit an index greater than 2, whereas 90\% of flare-productive regions exhibit an index greater than 2. Flare-quiet regions exhibit a high temporal variance (i.e, the index fluctuates between high and low values), whereas flare-productive regions maintain an index greater than 2 for several days. This shows the importance of including the temporal evolution of active regions in flare prediction studies, and highlights the potential of a 2-3 day prediction window for space weather applications.

\end{abstract}

\keywords{Sun, flares --- 
magnetic fields --- Photosphere}



\section{Introduction} \label{sec:intro}

     The localized dynamics of the magnetic field, rooted in the photosphere, gives rise to the formation, evolution, and decay of active regions. Flux emergence at the photopshere and the concentration of these magnetic fields lead to the formation of active regions, whereas the diffusion of these fields via supergranular buffeting results in their decay \citep{McAteer01}. The strong coupling between the magnetic field and velocity field is governed by the high magnetic Reynolds number in the fully-developed turbulent photosphere. Although all active regions are considered to be the manifestations of this large magnetic Reynolds number, not all active regions show the same level of solar activity. A few active regions tend to produce multiple, large flares, while many active regions evolve through their lifetimes without showing any considerable flare activity \citep{Bloom12}. This means that certain types of magnetic configuration may be more important in flare production \citep{McAteer05, Schrijver01, Al-Ghraibah01}. Detecting and understanding these differences in flare-productive active regions is essential to realize how and when solar flares occur \citep{Barnes01}.
     
     One common method used to study the magnetic energy distribution in active regions is the magnetic power spectrum. As introduced in detail by \citet{Abr05}, the magnetic power spectrum is generally obtained from a plot of energy $(E)$ against spatial wavenumber $(k)$. \citet {McAteer13} showed that the magnetic  power spectrum describes the distribution of magnetic energy content across all lengths scales present in an active region and can be separated into three ranges: the injective range, the inertial range, and the dissipation range. The injective range occurs at the large scale structures (i.e., small wavenumbers), where the energy is injected into the system. The inertial range occurs at intermediate values of scales and wavenumbers, where energy cascades down the structures. It is denoted by $E(k) \propto k^{-\alpha} $, where $\alpha$ is known as power index. The dissipation range occurs at small scales, where energy can be released from the system, usually in the forming of heat. 
     
     The power index, $\alpha$, describes the rate of energy transfer in the inertial range. Different power index values indicate the presence of different types of energy transfer mechanisms. A value of $5/3$ is known as a Kolmogorov type of spectrum which is generally observed for kinetic energy \citep{Kolmogorov41}. In the presence of magnetic field the power spectrum may take a form of $\alpha = 3/2$ \citep{Iroshnikov63, Kraichnan65}. Higher power index values (i.e., strongly non-Kolmogorov type steep spectra) indicate an intermittent or burst-like nature of energy dissipation. 
     
     The presence of such magnetic energy distribution may be one of the underlying causes of flare occurrence. Previous studies of a magnetic spectral analysis of active regions (\citet{Abr01} and references therein) showed that the presence of an intermittent energy dissipation is connected to the occurrence of solar flares. \citet{Abr05} analyzed 16 active regions with SoHO/MDI line-of-sight magnetograms and showed that active regions with significant flare activity possess a steeper non-Kolmogorov type power spectrum with $\alpha > 2$, while flare-quiet regions exhibit a Kolmogorov type spectrum. \citet{Hewett01} showed that the fast growing region NOAA 10488, evolved to exhibit $\alpha > 2$ 24 hours before the onset of multiple large solar flares. \citet{Abr10} expanded on this by expanding to a much larger scale study of 217 active regions, showing that, for regions which produced one flare, the power index and the total spectrum energy scaled with solar flare productivity. They also note that both emerging and decaying active regions tended to exhibit a lower index (closer to 5/3) compared to mature active regions (where the index tended to be 2 or steeper). This work was based on longitudinal magnetic field data, therefore was limited to $\pm$20 degrees from disk center. The Heliospheric and Magnetic Imager (HMI) onboard the Solar Dynamics Observatory (SDO) now overcomes this by providing full vector field data. Here we present a study using high-resolution SDO / HMI vector magnetograms of 53 HARPs (HMI active region patches) as they rotate across the solar disk, for a total of 2232 images, and including both flare-productive and non-flaring regions. The power index is calculated for each region to determine its relationship with flare productivity. Section \ref{sec:data} describes the data used for this analysis. Section \ref{sec:magpowind} shows the method used for the calculation of the power index. The flare productivity is defined in section \ref{sec:flareprod}, and the results are discussed in section \ref{disc}.

\section{Data} \label{sec:data}

SDO / HMI magnetograms provide the full vector field across the entire solar disk in a regular synoptic manner. These data are prepped according to \citet{Scherrer01} with the additional inversion as described in \citet{Bobra01} and \citet{Centeno01}. The resulting resolution for each image is 0.5 arcsec per pixel. Since SDO / HMI magnetograms provide magnetic field information in all three components, the vector magnitude is considered for magnetic power spectral calculations in these analysis. Each day, 6-7 images of each HARPS regions are extracted and analyzed.

\section{Magnetic Power Index} \label{sec:magpowind}
The magnetic power index is defined as the negative of the slope of the inertial range of the power spectrum.  The power spectrum is defined as \citep*{Mon75ya} :
\[
E(k) = \int\limits_{\left|\textbf{k}\right|=k}\,F(\textbf{k})\,dS(\textbf{k})\
\]
where $F(\textbf{k})$ is the Energy spectrum of 2D function which is obtained as \citep*{Mon75ya}:
\[ 
F(\textbf{k}) =\left|U(\textbf{k})\right|^{2}
\]
Here $U(\textbf{k})$ is the Fourier transform of an image and \textbf{k} is the wave vector. 
\[
k = \sqrt{k_x^2 + k_y^2}
\]
where $k_x$ and $k_y$ are the spatial frequency values along the respective axes and $dS(\textbf{k})$ is the length element of the circle with radius $k$ that falls within the region bounded by $k_x$ and $k_y$. To compute the magnetic power spectrum we used the technique as described in detail in \citet{Abr01}. All $F(k_x, k_y)$ within an annulus are added and this summation is divided by the number of points within the annulus. This average is then multiplied by the length element of the circle with radius $k$ that lies within the Fourier plane. This ensures that the integration is carried out over an appropriate length at high $k_x$ and $k_y$. This is a vital stage as it allows for quiet-Sun contamination to be removed (Section \ref{subsec:imgcorr})

\begin{figure}    
\figurenum{1}
\gridline{\fig{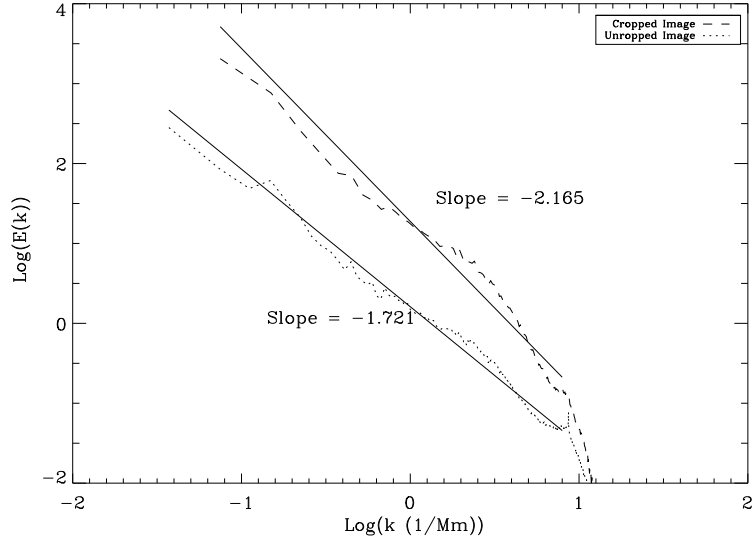}{0.6\textwidth}{}}
\caption{Dotted line indicates the spectrum for the original mangetogram of HARP 1028 (NOAA 11339). Dashed line shows the spectrum for the carefully cropped magnetogram.\label{fig:fig1}}
\end{figure}

\begin{figure}    
\figurenum{2}   
\gridline{{\fig{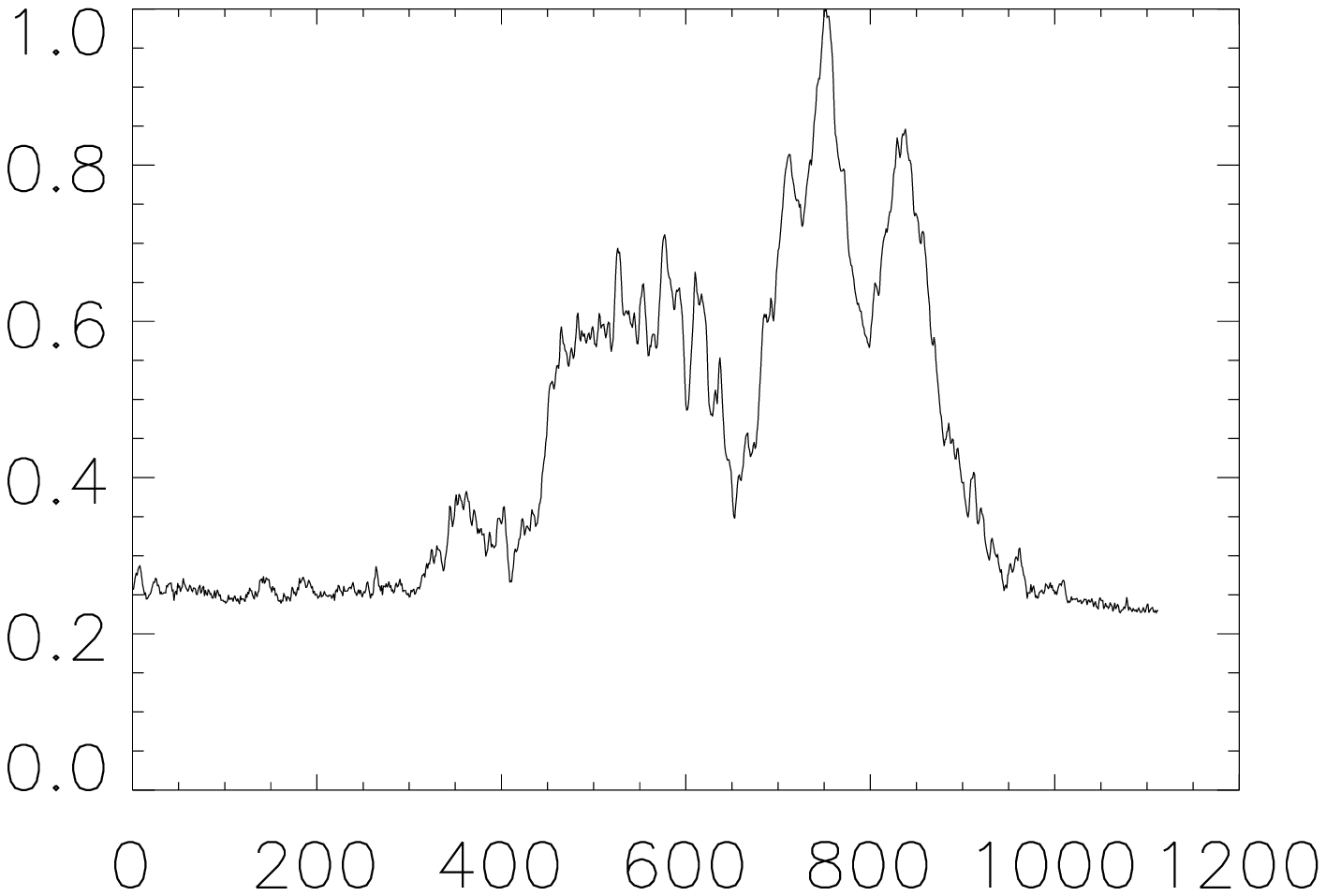}{0.45\textwidth}{(a) xvalues}}
          {\fig{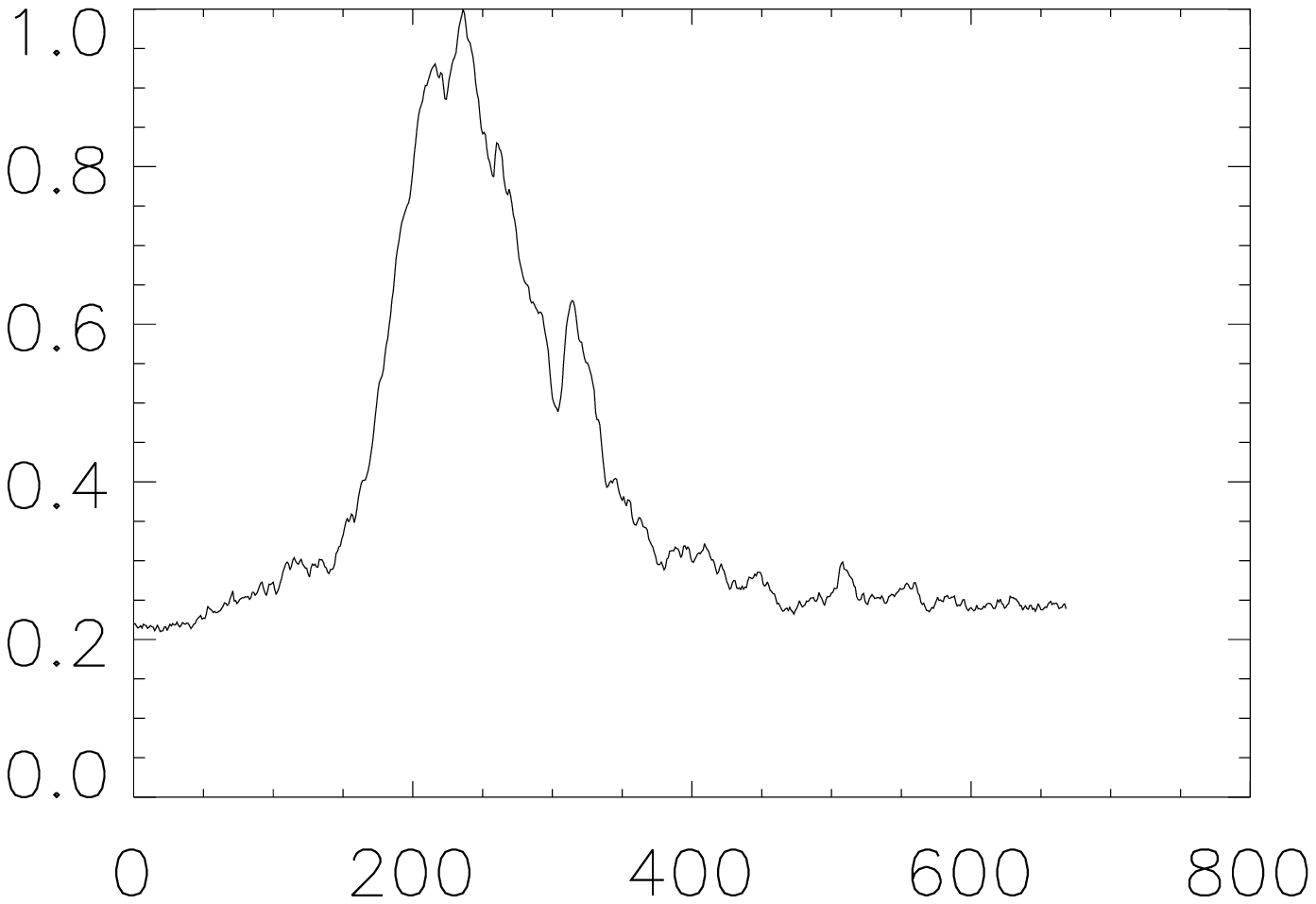}{0.45\textwidth}{(b) yvalues}}
          }
\gridline{{\fig{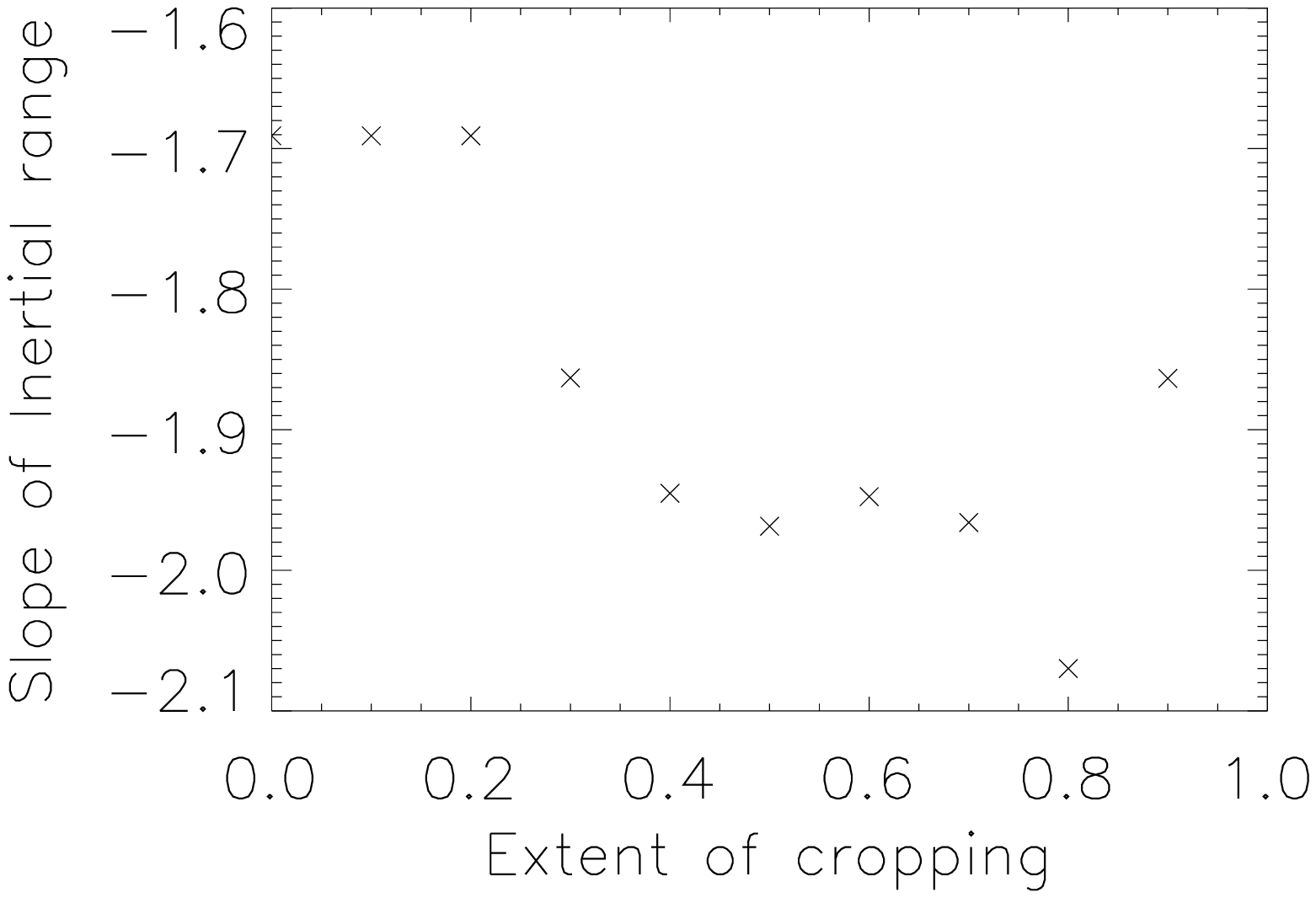}{0.45\textwidth}{(c) Slope variation in x direction} }
          {\fig{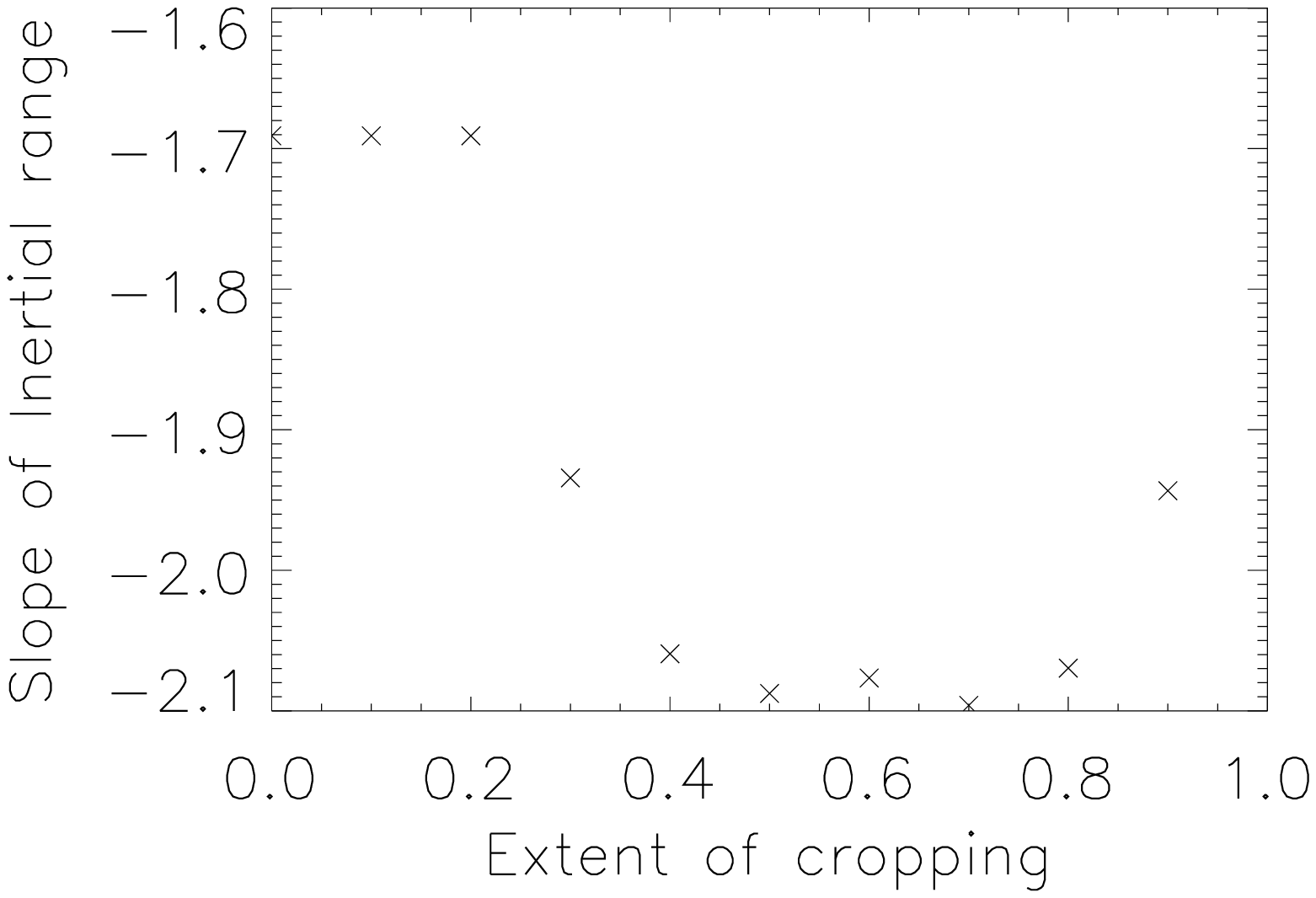}{0.45\textwidth}{(d) Slope variation in y direction} }
          }
\caption{Upper panel shows xvalues and yvalues for the original magnetogram of HARP 1028 (NOAA 11339). Lower panel shows the variation of slope values as the image is cropped at different threshold values. First three slope values are same since all xvalues and yvalues are above 0.2. At the threshold value of 0.3, we see a sharp drop in the slope. This happens due to the omission of the quiet Sun portion around the active region.\label{fig:fig2}}
\end{figure}

\begin{figure}  
\figurenum{3}
\gridline{\fig{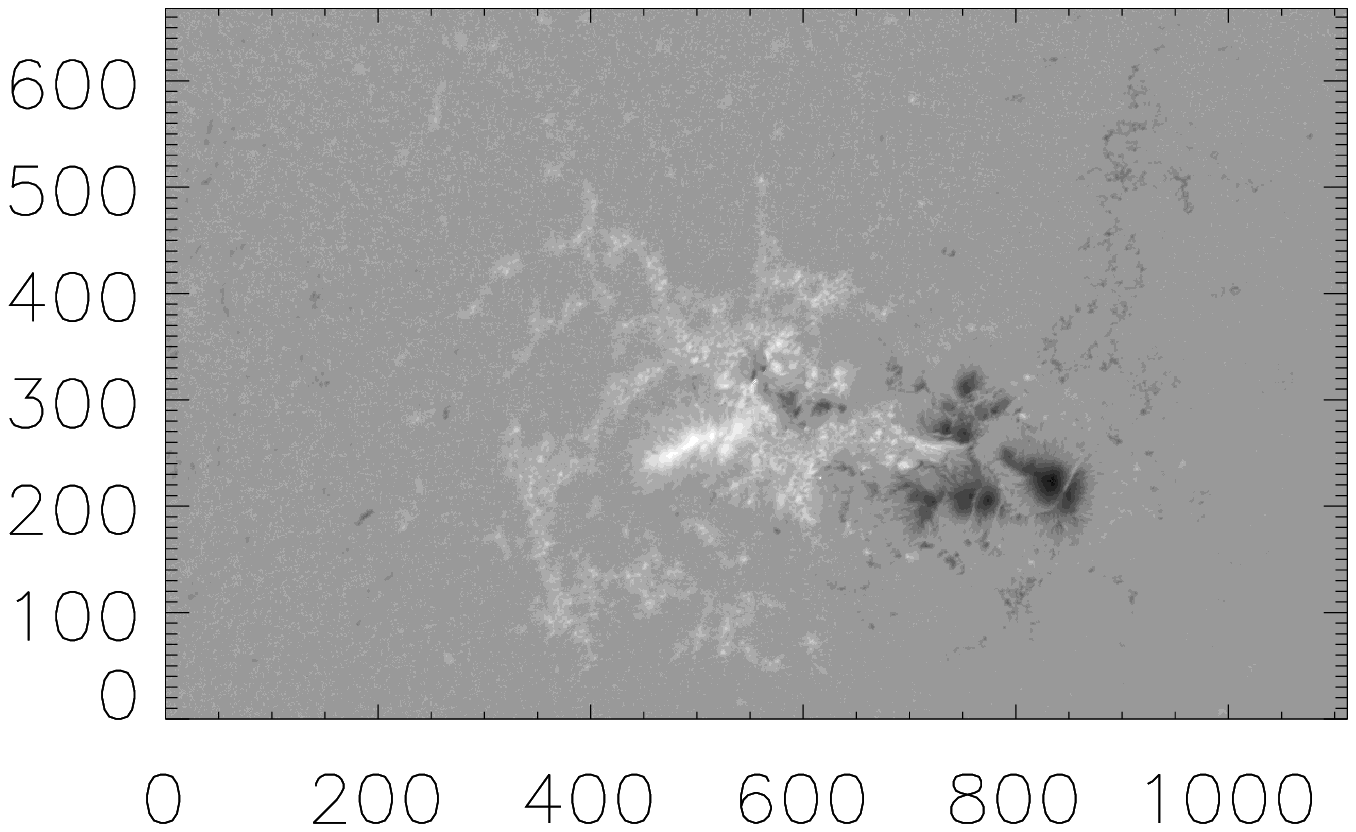}{0.45\textwidth}{}
          \fig{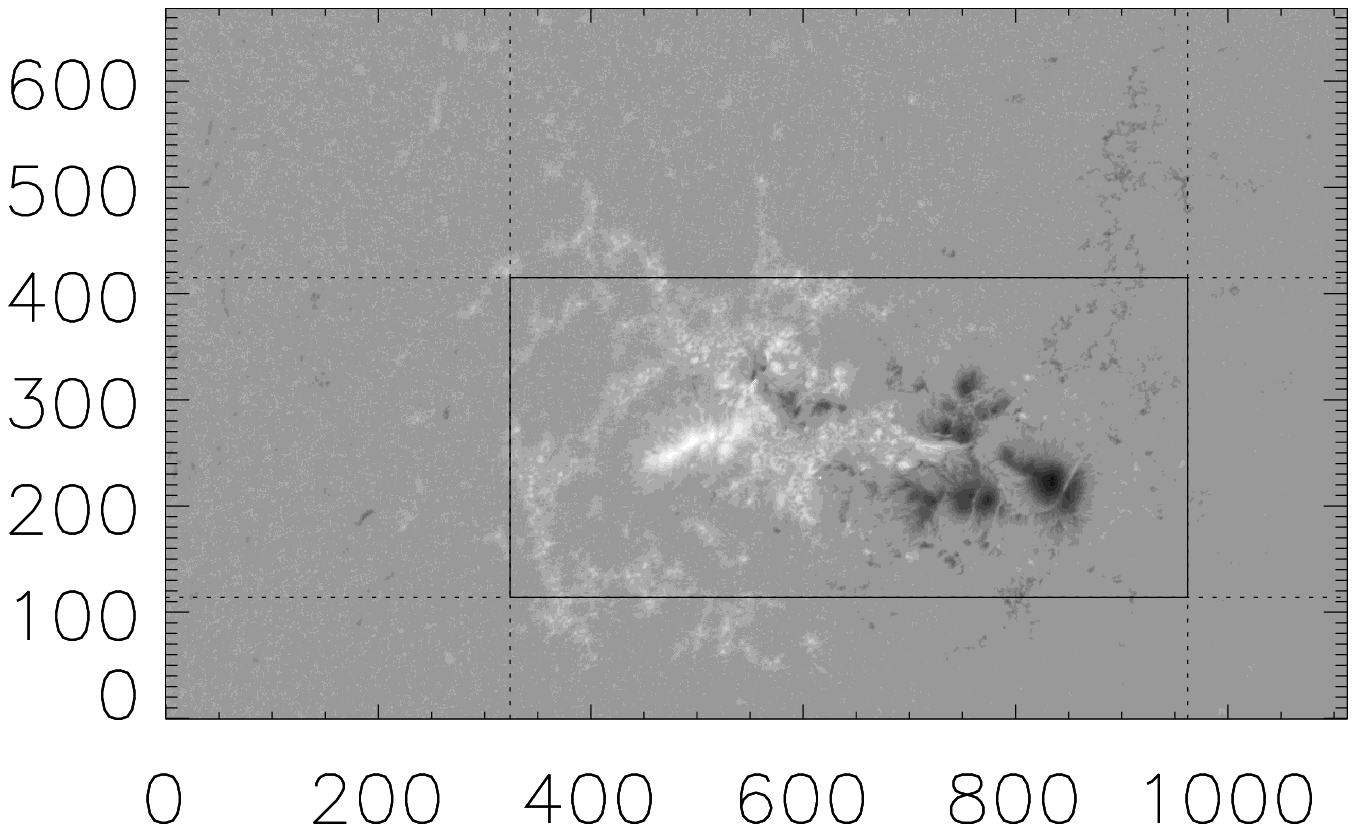}{0.45\textwidth}{}
          }
\caption{Left: Original uncropped image of HARP 1028(NOAA 11339). Right: Cropped image of HARP 1028 (NOAA 11339).\label{fig:fig3}}
\end{figure} 

\begin{figure}    
\figurenum{4}
\gridline{\fig{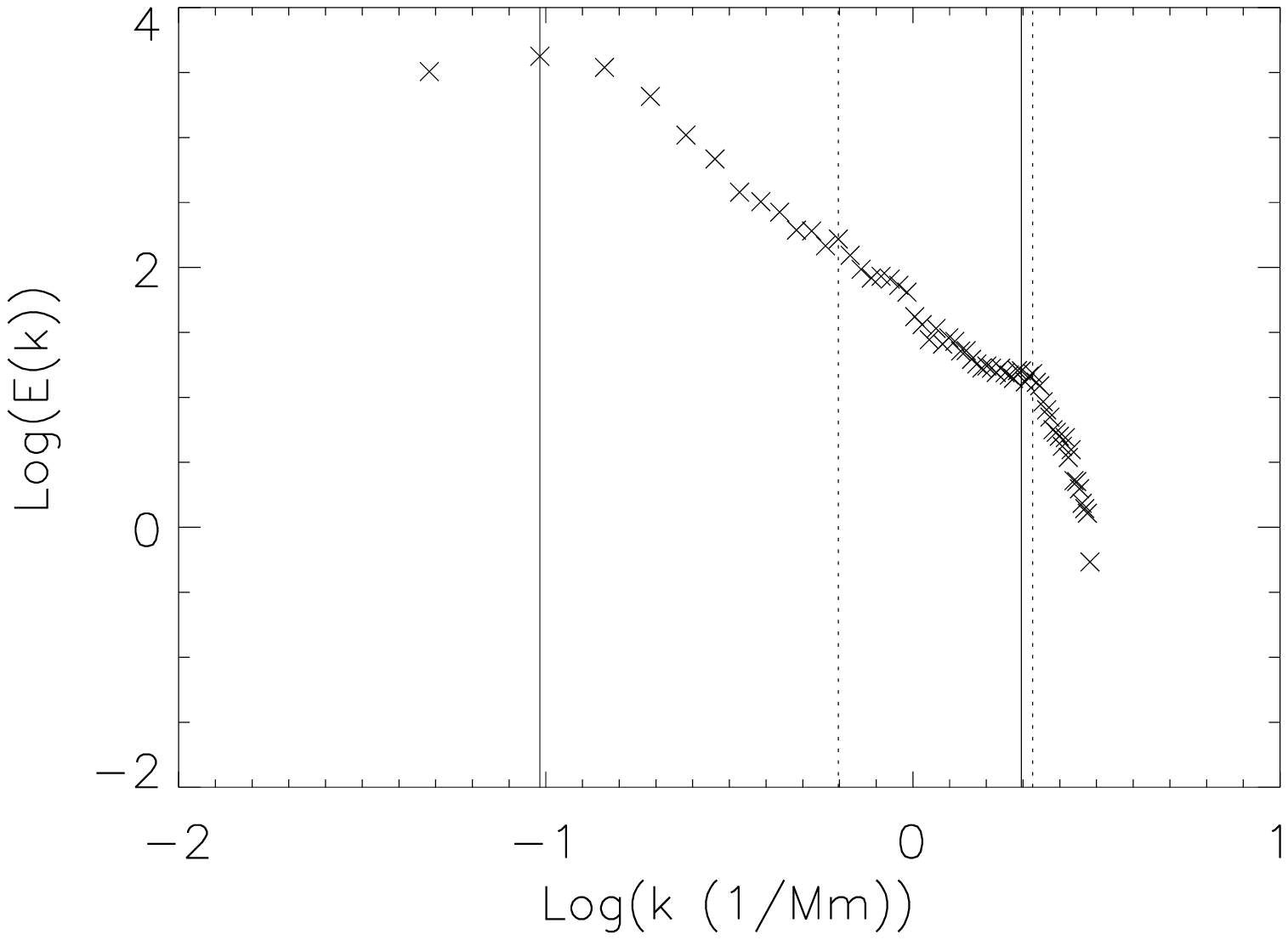}{0.6\textwidth}{}}
\caption{Power Spectrum showing the determination of the inertial range for spectral fitting. Dotted vertical lines show an inertial range of 3-10 Mm. Solid vertical lines show the inertial range based on the spectral discontinuities. \label{fig:fig4}}
\end{figure}

\subsection{Image Correction} \label{subsec:imgcorr}

      In general, a HARP image includes some information from both active region and the surrounding quiet Sun. The quiet Sun is mostly small scale features, and contains its own characteristic field distribution \citep{Conlon01,McAteer02}. Hence any active region image that contains large areas of quiet Sun will exhibit excess power at large $k$, lowering the overall index value. The distortion of the resulting spectrum is shown in Fig.\ref{fig:fig1}. Hence, the optimum cropping of each image is an essential first step in the data analysis. The image is cropped at various sizes to determine the optimum window that excludes most quiet Sun, while including most active region. The following approach is used to remove the quiet sun portion around the active region. Magnetic field values at each pixel are added along columns and normalized (xvalues) (Fig. \ref{fig:fig2}(a)) and along rows and normalized (yvalues) (Fig. \ref{fig:fig2}(b)). Peaks show the regions of high magnetic values whereas a flat portion indicates the low value quiet Sun portion in the image along x and y direction. These values are normalized so that the method can be applied to all the images. Since the vector magnitude is considered, these values are positive. 

   The image is cropped along the x direction at 10 different proportions, keeping the number of pixels in y direction constant. The size of the image in the x direction is chosen such that the first and last pixels are the points where xvalues are above the threshold. The threshold values are 10 equally spaced points between 0.0 to 0.9 (Fig. \ref{fig:fig2}(a)). Power spectra are obtained for these cropped images and the slope calculations are performed. Figure \ref{fig:fig2}(c) shows how slope values vary as the images is cropped more and more as the threshold value is increased.
   
Similarly, the image is cropped in the y direction at 10 different proportions, keeping the original number of pixels in the x direction constant. Figure \ref{fig:fig2}(d) shows the slope variation with the cropping of the image.

   From Figure \ref{fig:fig2}(c) and Figure \ref{fig:fig2}(d), the first three slope values are same, since xvalues and yvalues lie above 0.2 and therefore nothing has been cropped out. This provides the initial index for the entire image (active region and quiet Sun). At higher thresholds, there is a considerable drop in the slope value. This is because most of the flat portion in figure \ref{fig:fig2}(a) and figure \ref{fig:fig2}(b) (i.e., the quiet Sun portion) has been cropped out. At higher threshold values, the index drops off once more as only the largest values of magnetic field (the central portions to the active region) as included. In figure \ref{fig:fig2}(c) and figure \ref{fig:fig2}(d), a threshold value of 0.3 provides the desired result of removing the most quiet Sun while ensuring all the active region is included in the analysis. From this threshold value, the first and the last pixels in both x and y axes of the image are obtained (Fig. \ref{fig:fig3}). In a few cases where this method doesn't work due to particular arrangements of spots (e.g., when two NOAA regions appear in one HARP), the image is cropped manually.  
      
\subsection{Inertial Range} \label{subsec:inrng}

The power index should only be computed from the inertial range of the power spectrum, therefore the selection of the correct start and end of the inertial range is very important. In previous power spectral studies performed on SoHO / MDI and ground based data, the inertial range is considered to fall within 3-10 Mm \citep{Abr05, Abr10, Hewett01}. The lower limit of the inertial range is considered to lie at the wavenumber $k = 2\pi/10$. But this value may not define the lower limit of the inertial range for every active region. As an example, Figure \ref{fig:fig4} shows the magnetic power spectrum for NOAA 10488. The dotted vertical lines show this conventional, 3-10 Mm, inertial range whereas solid vertical lines show the larger inertial range based on spectral discontinuities. The lower limit for the inertial range takes the value much less than $2\pi/10$. The upper limit, $2\pi/3$ is also not a special value as it depends on the resolution of the image. Indeed the size of the inertial range may itself be an important parameter in the study of active regions magnetic structure \citep{McAteer16b}. 

In this study, we define the inertial range of the power spectrum as that part of the spectrum that lies between two spectral discontinuities. The first spectral discontinuity separates the injection range from the inertial range. This discontinuity occurs at the wave number of maximum energy and sets the lower limit of the inertial range. The second spectral discontinuity separates the inertial range from the dissipation range. The dissipation range is much steeper than the inertial range. The discontinuity provides the upper limit of the inertial range. In this study, where the inertial range is determined on an individual basis, based on these spectral discontinuities, the inertial range occupies (and hence the spectral fitting is carried out over) on average approximately 65\% of the spectrum. 

\begin{figure}    
\figurenum{5}
\gridline{\fig{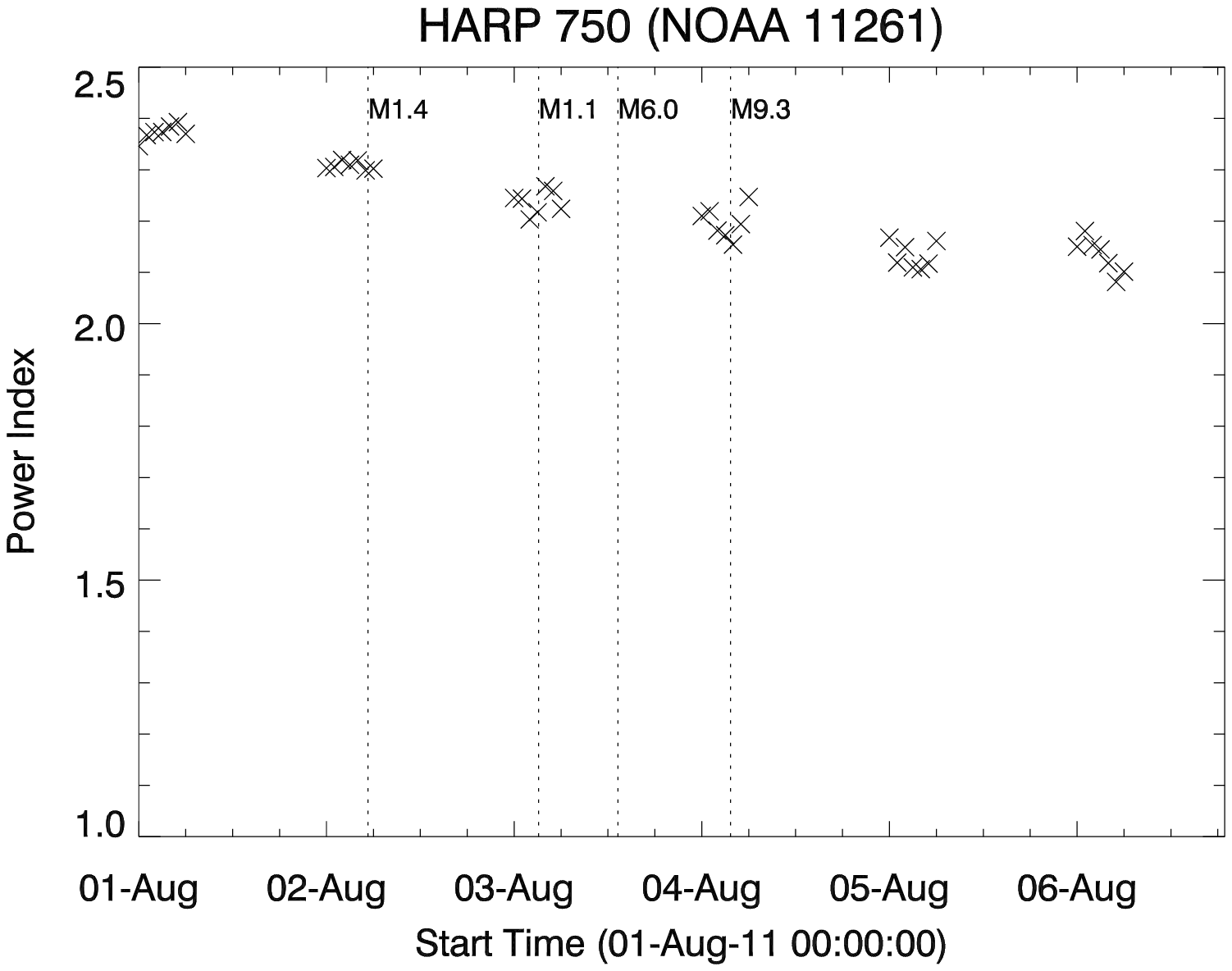}{0.45\textwidth}{\label{fig:fig5:HARP750}}
          \fig{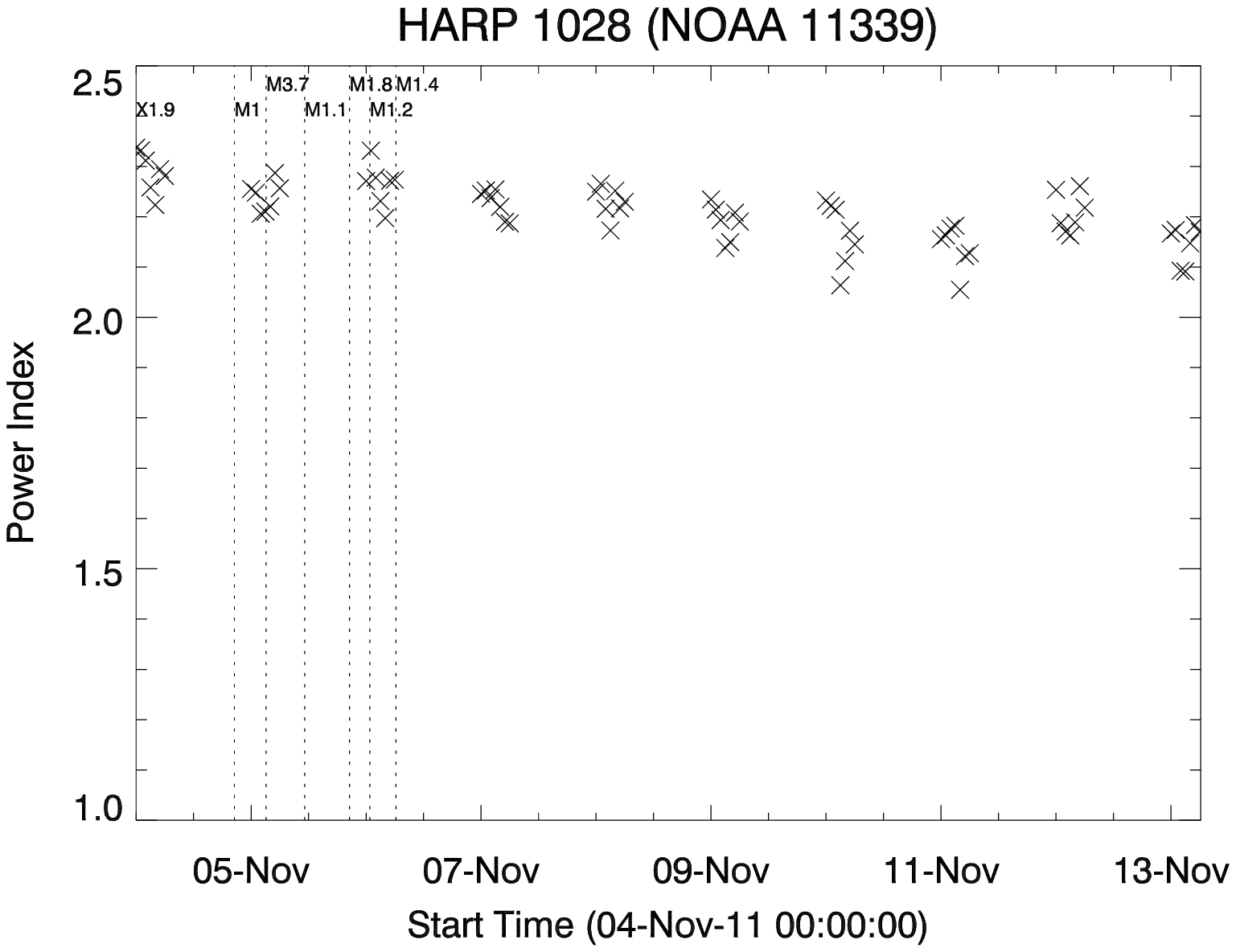}{0.45\textwidth}{\label{fig:fig5:HARP1028}}
          }
\gridline{\fig{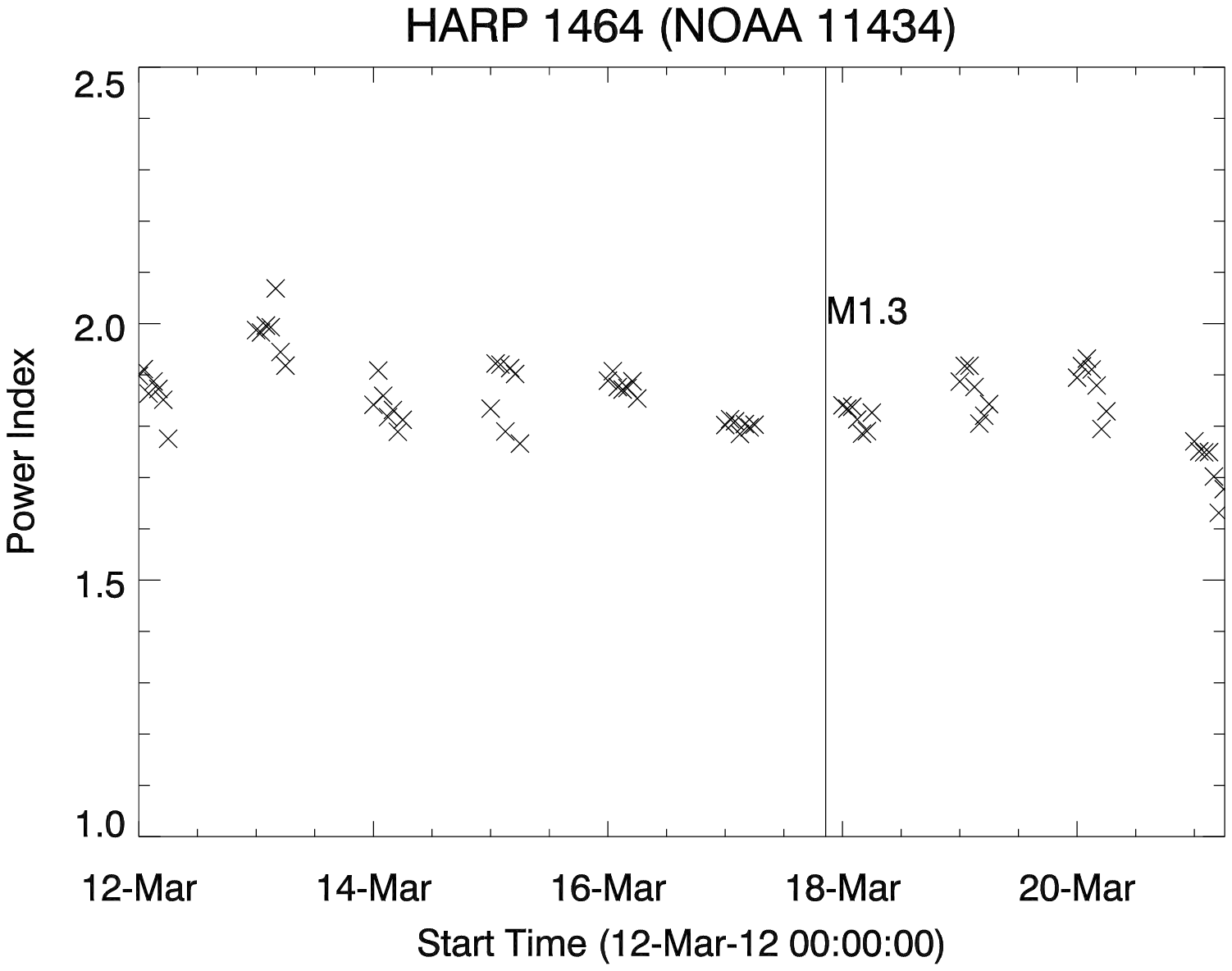}{0.45\textwidth}{\label{fig:fig5:HARP1464}}
          \fig{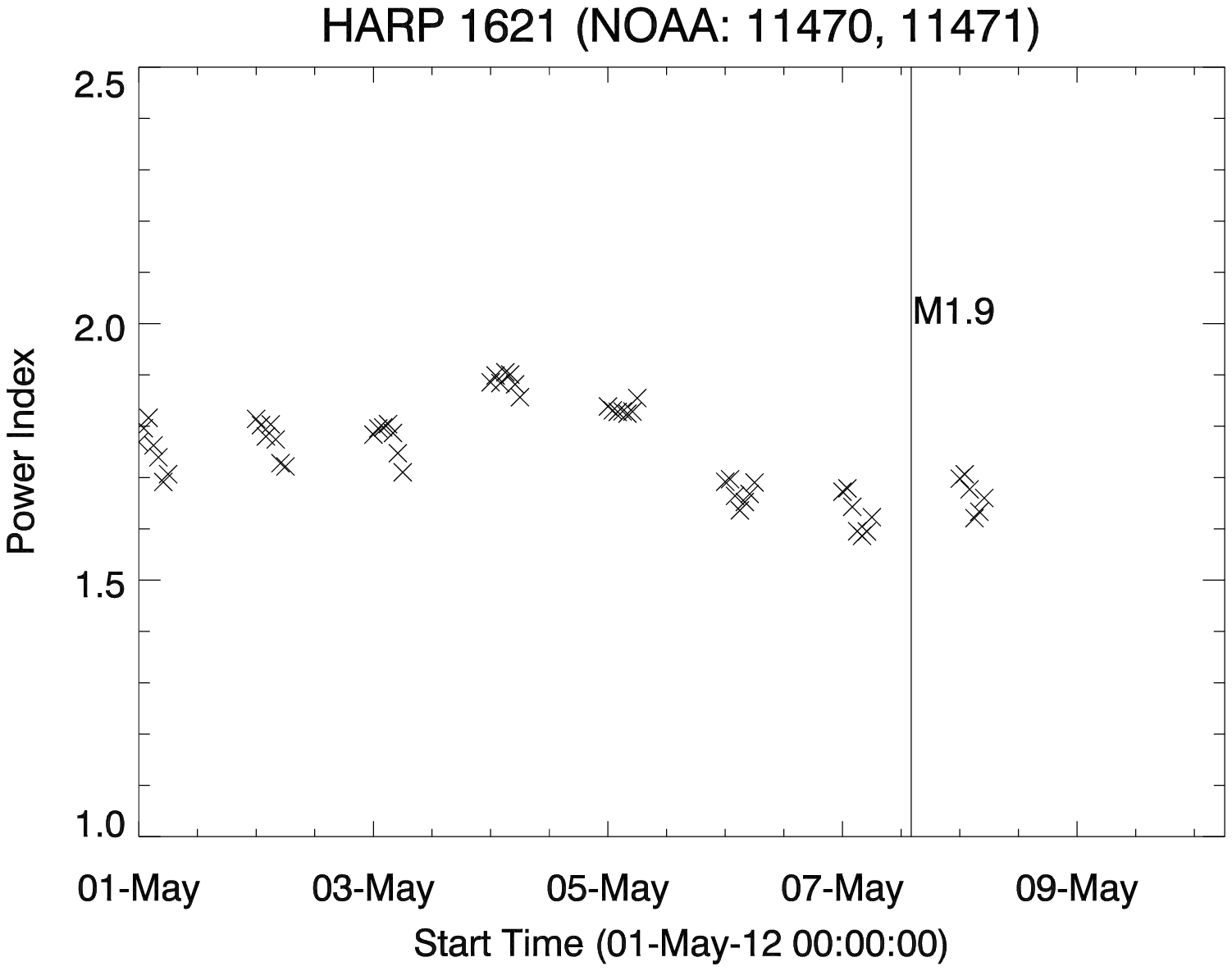}{0.45\textwidth}{\label{fig:fig5:HARP1621}}
          }
\gridline{\fig{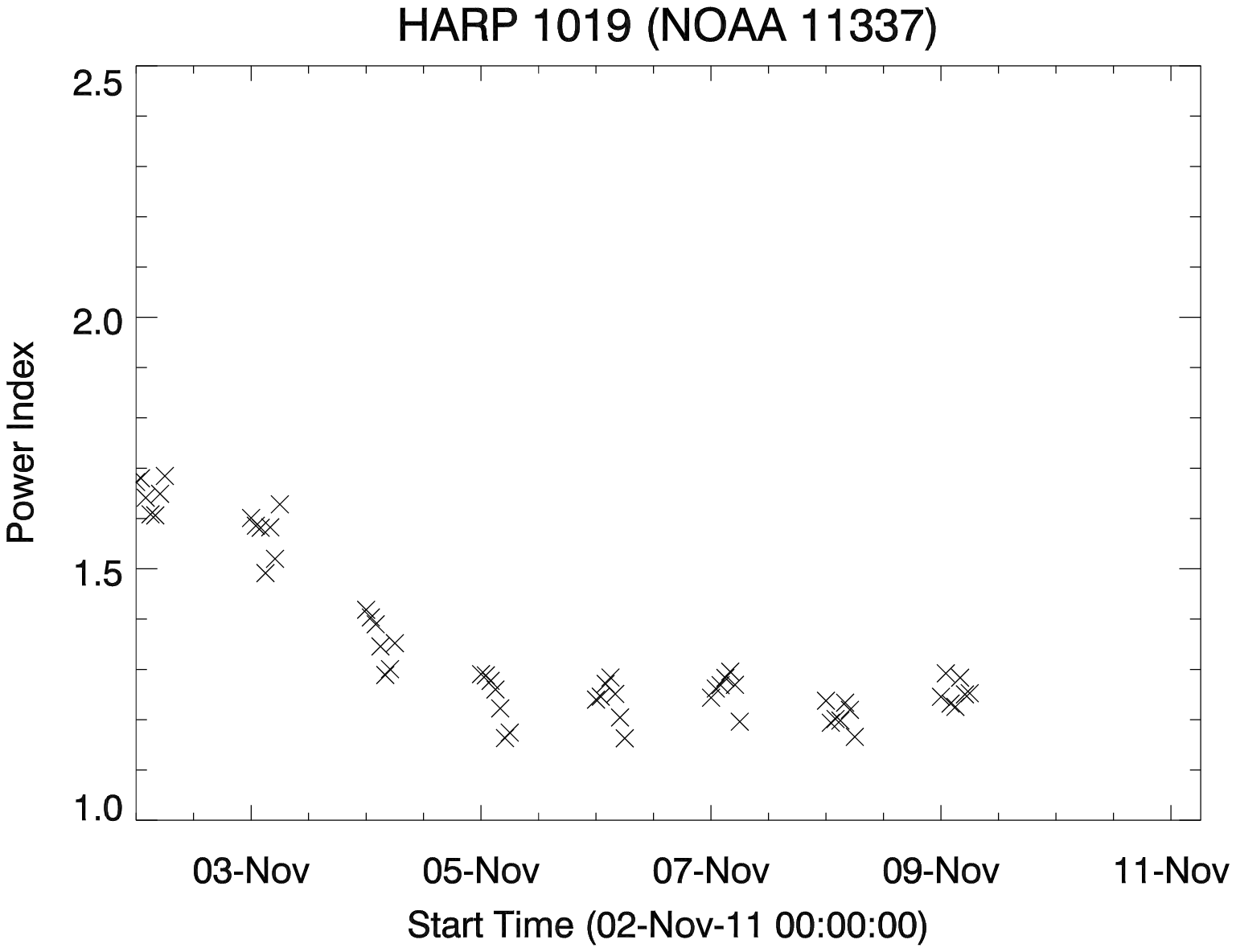}{0.45\textwidth}{\label{fig:fig5:HARP1019}}
          \fig{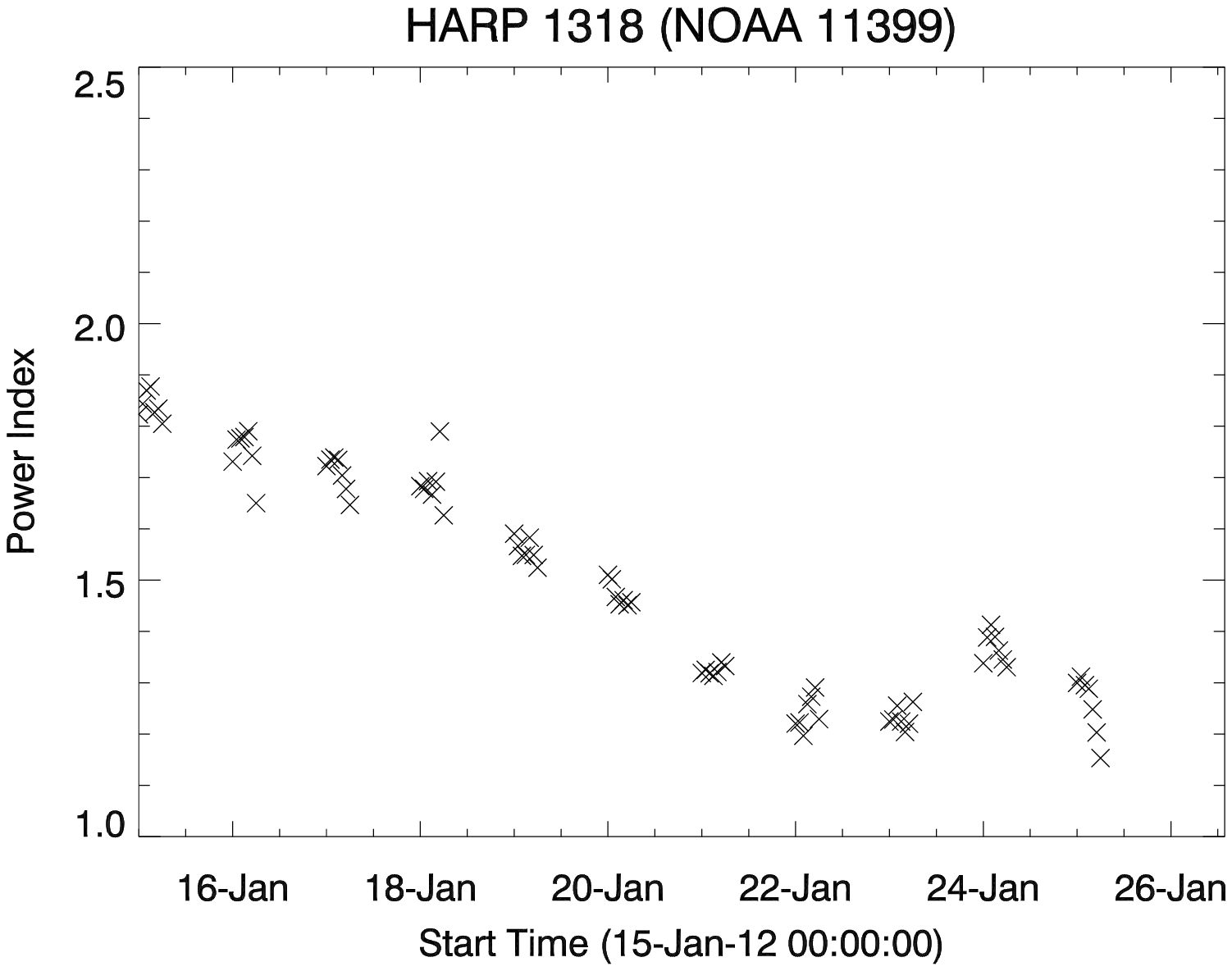}{0.45\textwidth}{\label{fig:fig5:HARP1318}}
          }
\caption{Upper Panel: Power index evolution for HARP 750 and HARP 1028. These flare-productive active regions produced at least one flare of intensity greater than M5. Middle Pane: Power index evolution for HARP 1464 and HARP 1621. These flare-productive active regions produced flares of intensity between M1 and M5. Lower Panel: Power index evolution for HARP 1019 and HARP 1318. These active regions showed no flare productivity. Vertical lines in the graphs indicate the occurrence of flares with intensity greater than M1.\label{fig:fig5}}
\end{figure}   

\section{Flare Productivity} \label{sec:flareprod}

The data studies here consist of time series of 53 HARPs selected from August 2011 to July 2012. This data contains active regions that exhibit a wide range of flare activity. In order to study the dependence of power index on flare activity, these HARPs were categorized based on the sizes and numbers of flares they produced. If a region produced no flares or flares of intensity less than C1, it was designated a flare-quiet active region. If a region produced at least one flare of intensity greater than M1 then it was considered to be a region of high flare productivity. 25 HARPs showed no flare activity whereas 28 HARPs were highly flare productive. Out of these 28 HARPs, 19 HARPs produced flares with intensity less than M5, whereas 9 HARPs produced at least one flare with intensity greater than M5.

\begin{figure}    
\figurenum{6}
\gridline{\fig{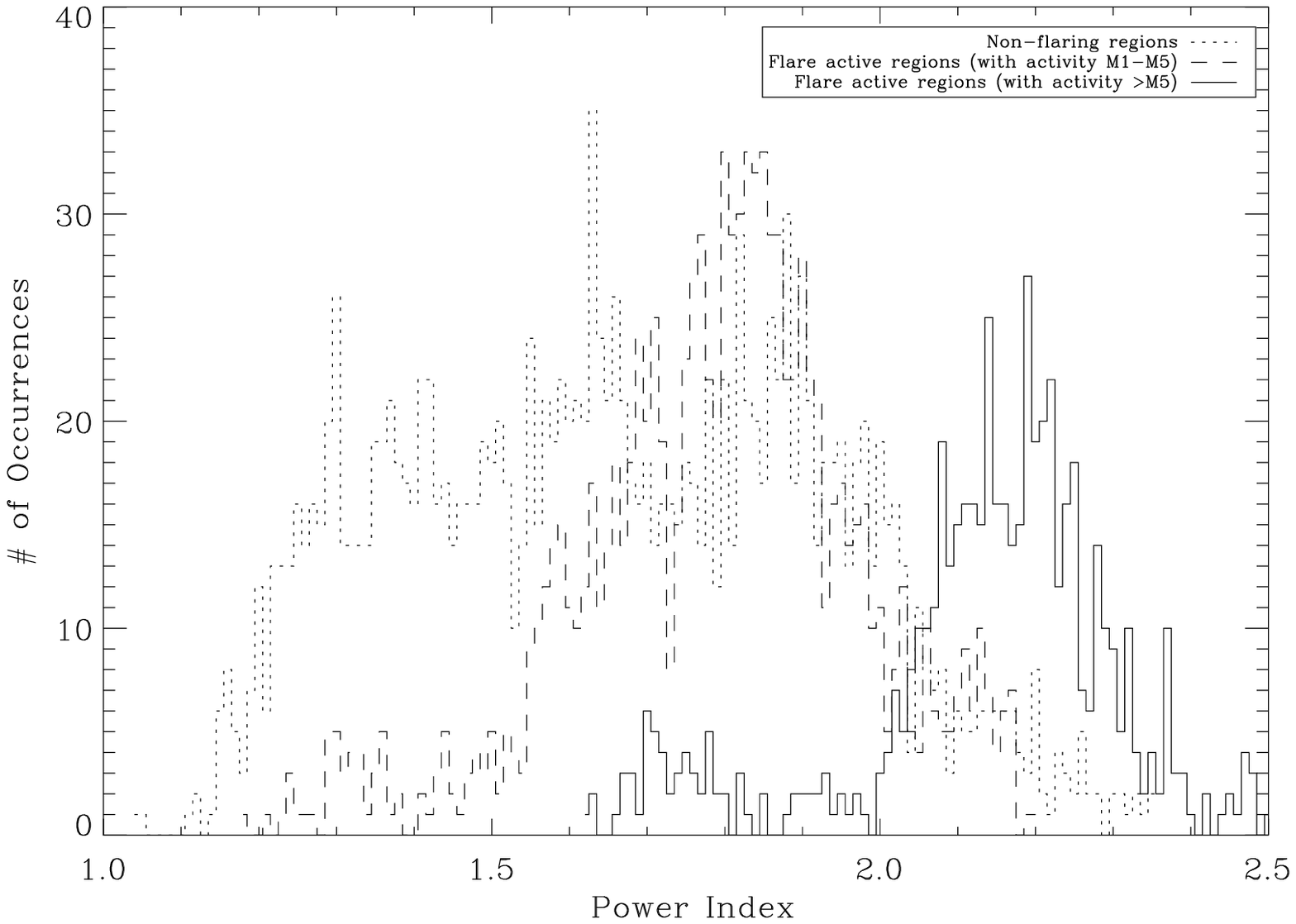}{0.6\textwidth}{}}   
\caption{Histograms of power index values for all 53 HARPs. The dotted line indicates a histogram for 25 flare-quiet active regions. The dashed line shows a histogram for 19 flare-productive active regions that produced flares of intensity between M1 and M5. The solid line indicates a histogram for 9 flare-productive active regions that produced at least one flare with intensity greater than M5. \label{fig:fig6}}
\end{figure}

\begin{figure}
\figurenum{7}
\gridline{\fig{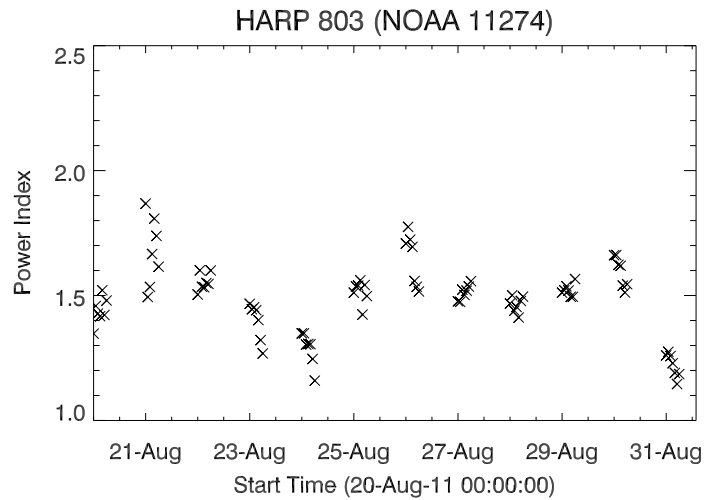}{0.6\textwidth}{}}
\caption{Power index evolution of a non-flaring active region HARP 803(NOAA 11274). Flare quiet regions tend to show much variation in the power index values.\label{fig:fig7}}
\end{figure}

\begin{figure}    
\figurenum{8}
\gridline{\fig{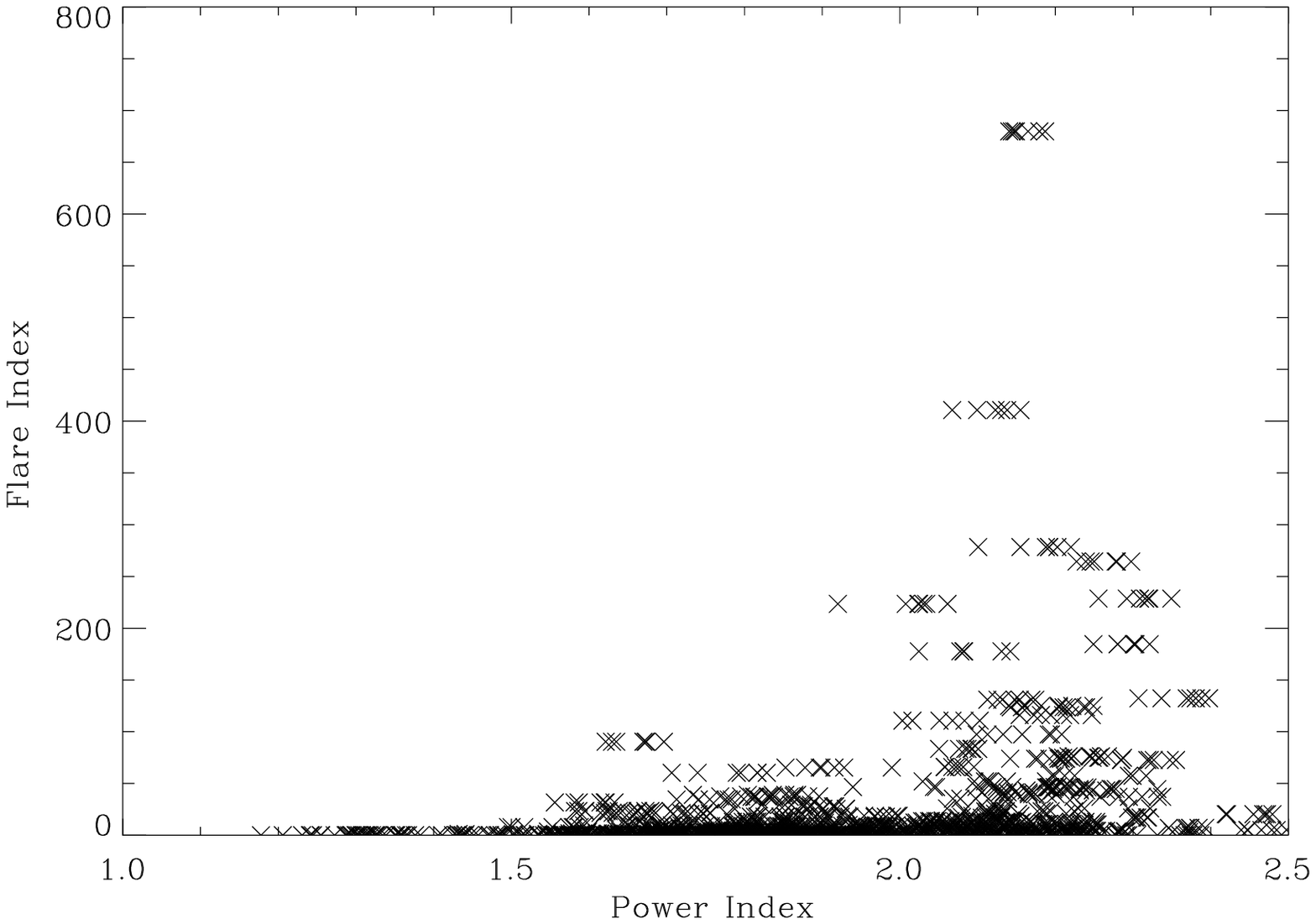}{0.6\textwidth}{}}   
\caption{The flare index is plotted against the power index for 28 flare-productive active regions. The flare index increases with an increase in the power index and achieves the maximum at round 2.2. \label{fig:fig8}}
\end{figure} 

Figure \ref{fig:fig5} shows the power index evolution for six example datasets. The upper panel shows the power index evolution, over the timerange of all available data, for two flare-productive active regions that produced at least one flare of intensity greater than M5. In every 24 hour period, a cluster of 6-7 images are studied, allowing for us to study both daily changes (the preferred timescale for space weather predictions) and short time changes, without being overawed with data volume. The power index values mostly lie between 2 to 2.5. The middle panel shows the power index evolution for two flare-productive active regions that produced flares of magnitude M1 - M5. In these cases, the power index takes a value between 1.5 to 2. The lower panel shows the results for the two flare quiet active regions. Here the power index lies mostly below 1.5. 

In addition to exhibiting, on average, a lower value for the power index, flare-quiet regions tends to show much more variability in the temporal evolution of the index. Figure \ref{fig:fig5} (bottom) shows one such trend - where the index tends to drop off over several days. A second trend is shown in Figure~\ref{fig:fig7}, where NOAA 11274 shows periods when the power index rises to 1. However, these periods of time are short. In contrast, there is no sudden spike in the index value for any of the flare-productive regions as they track across the solar disk. Instead flare productive regions exhibit a large power index for many days, both before and after they produce solar flares. This suggests that the flare potential of any active region could be determined a few days in advance of the production of a large solar flare and the occurrence of solar flares do not reduce the future flare potential. 

 Figure \ref{fig:fig6} shows the results in the form of histograms for all 53 HARPs. The range of index values ($1.0-2.5$) is similar to that in previous studies \citep{Abr10}. The dotted histogram shows the power index values taken by flare-quiet active regions. The power index distribution for these flare-quiet regions is quite broad and peaks around 1.6. The dashed histogram indicates the results for flare-productive active regions that produced flares of magnitude M1 - M5. This distribution is much narrow, with fewer values below 1.5, and peaks around 1.8. However the high-value tail does overlap with the flare-quiet distribution. Finally, the solid histogram shows the power index values taken by flare-productive active regions that produced at least one flare of intensity greater than M5. In this case, the power index mostly lie above 2 with a maximum at around 2.2.   

The above classification of HARPs was performed based on the maximum intensity of any flare from the region. However, the flare productivity can be measured more carefully in terms of the Soft X-ray (SXR) flare index. This index was first introduced by \citet{Antalova96}. To compute the SXR flare index, flares of classes B, C, M, and X are weighted as 0.1, 1, 10, and 100 respectively. The flare index is then computed as follows \citep{Abr05}:
\[
          A = (100S^{(X)}+10S^{(M)}+1.0S^{(C)}+0.1S^{(B)})/\tau
\]

Here S is the sum of GOES peak intensities of a particular class. (i.e. X, M, C, B classes and intensities are denoted by indices from 1.0 to 9.9 following these classes). and $\tau$ is the time measured in days. Although sensitive to properties of the electron beam in any particular event \citep{Reep13}, the GOES magnitude remains the best estimated of flare size.

The flare index was computed on a daily basis for all 28 flare-productive active regions. Figure \ref{fig:fig8} shows the dependence of the flare index on the power index values. The highest flare index values are only associated with power index values greater than 2. As in Figure \ref{fig:fig6}, the power index distribution exhibits a maximum around 2.2.

\section{Discussion} \label{disc}

A magnetic power spectral analysis was performed on 53 HARPs as they travelled across the solar disk. A Fourier transform was performed on the full vector field to obtain the magnetic power index for each dataset.The power spectrum and thereby the power index is affected by the presence of low-value quiet sun portion around the active region. An optimum spectrum is obtained by excluding the quiet sun region around the active region. The inertial range is determined on a case-by-case basis, and often lies outside the conventional 3-10 Mm region of the spectrum.
            
The power index is shown to be significantly larger for active regions with larger flare productivity. The histograms for flare-quiet region and highly flare-productive active regions are well-separated. The power index for flare-quiet active regions peaks around 1.6 which is close to the Kolmogorov value. The power index values for highly flare-productive active regions tend to be greater than 2 which represents a non-Kolmogorov type of steep spectra. This confirms the results of \citet{Abr10} can be extended to full vector field data and provides further evidence to suggest the presence of different types of underlying energy dissipation mechanism in flare productive regions. The Kolmogorov type spectra indicate the smooth energy dissipation whereas the non-Kolmogorov type steep spectra shows that the energy dissipation is much more intermittent.
      
A power index of 2 seems to define the boundary for flare-productive regions for longitudinal data obtain close to disk center \citep {Abr01, Abr05}. We confirm that when the full vector field is studied across the solar disk, for both flare-quiet and flare productive regions, flare-productive regions exhibit a higher power index (i.e., only 10\% of all flare-productive regions have power index less than 2) and flare-quiet regions exhibit a smaller power index (i.e., only 12\% of all quiet regions have power index greater than 2). Importantly we show for the first time that flare-productive regions tend to exhibit a small variance around a large value, whereas flare-quiet regions can exhibit sudden fluctuations between larger and smaller values. This suggests that flare-quiet regions can experience some sudden periods of rapid strong inertial decay from large sizescales to small sizescales, but this is sufficiently brief that the magnetic field can adjust to compensate without releasing energy in solar flares. In contrast, flare-productive regions maintain a steep, non-Kolmogorov spectrum for many days, the magnetic field becomes stressed as magnetic energy builds up at small size scales, and solar flares become the inevitable consequence.

For this study, only regions that produced either no flares or at least one flare of intensity greater than M1 were considered. Regions that produced only C-class flares were not considered. In addition there was no distinction made in relation to CME productivity. In future work it will be essential to analyze many more regions to provide these extra classifications. Another important improvisation that arises is the requirement to track active regions for long periods of time. The Sun limits this to a maximum of about 10 days, but simulations of active region evolution should be used to overcome this boundary. Determining the power index evolution of active regions from their formation to their decay is essential to discover to the stage in the active region evolution where the power index will be most relevant in determining the occurrence of solar flares \citep{Boucheron01}.

\acknowledgments

 The data used here was made possible by funding to NWRA from NASA/LWS contract NNH09CE72C (Dr. Graham Barnes, PI).
This work was funded by NASA Heliophysics GI NNH12CG10C and NSF Career 1255024.

\end{document}